\documentclass[12pt]{iopart}
\usepackage{lineno,hyperref}
\usepackage{amsfonts}
\usepackage{graphicx}

\begin{document}

\review{Quantum sensing of microwave electric fields based on Rydberg atoms}

\author{Jinpeng Yuan$^{1,2,\dag}$, Wenguang Yang$^{1,2,\dag}$, Mingyong Jing$^{1,2}$, Hao Zhang$^{1,2}$, Yuechun Jiao$^{1,2}$, Weibin Li$^3$, Linjie Zhang$^{1,2,4,*}$, Liantuan Xiao$^{1,2}$, and Suotang Jia$^{1,2}$}

\address{$^1$ State Key Laboratory of Quantum Optics and Quantum Optics Devices, Institute of Laser Spectroscopy, Shanxi University, 92 Wucheng Road, Taiyuan 030006, China}
\address{$^2$  Collaborative Innovation Center of Extreme Optics, Shanxi University, 92 Wucheng Road, Taiyuan 030006, China}
\address{$^3$ School of Physics and Astronomy, and Centre for the Mathematics and Theoretical Physics of Quantum Non-equilibrium Systems, University of Nottingham, Nottingham NG7 2RD, United Kingdom}
\address{$^4$ Hefei National Laboratory, Hefei 230088, China}

\ead{zlj@sxu.edu.cn}
\vspace{10pt}
\begin{indented}
	\item[]July 2023
	\item{$\dag$ These authors contributed equally to this work.}
	\item{$*$ Author to whom any correspondence should be addressed.}
\end{indented}

\begin{abstract}
Microwave electric field sensing is of importance for a wide range of applications in areas of remote sensing, radar astronomy and communications. Over the past decade, Rydberg atoms, owing to their exaggerated response to microwave electric fields, plentiful optional energy levels and integratable preparation methods, have been used in ultra-sensitive, wide broadband, traceable, stealthy microwave electric field sensing. This review first introduces the basic concept of quantum sensing, properties of Rydberg atoms and principles of quantum sensing of microwave electric fields with Rydberg atoms. Then an overview of this very active research direction is gradually expanded, covering progresses of sensitivity and bandwidth in Rydberg atoms based microwave sensing, superheterodyne quantum sensing with microwave-dressed Rydberg atoms, quantum-enhanced sensing of microwave electric field, recent advanced quantum measurement systems and approaches to further improve the performance of microwave electric field sensing. Finally, a brief outlook on future development directions is discussed.
\end{abstract}

\newpage
%
%
%
%
%
\tableofcontents
\maketitle

\section{Introduction}
\label{s1}
Microwave electric fields (MW E-fields) are defined as electromagnetic radiations with frequency ranging from 300 MHz to 300 GHz corresponding to wavelengths ranging from one meter to one millimeter. Properties, such as easy to focus, high directivity and large data capacity, make them efficiently transmit in unobstructed line-of-sight free space, and have a wide range of applications in remote sensing, radar astronomy and communications \cite{Hitchcock2004}.

\begin{figure}[ht!]
	\includegraphics[width=13.0cm]{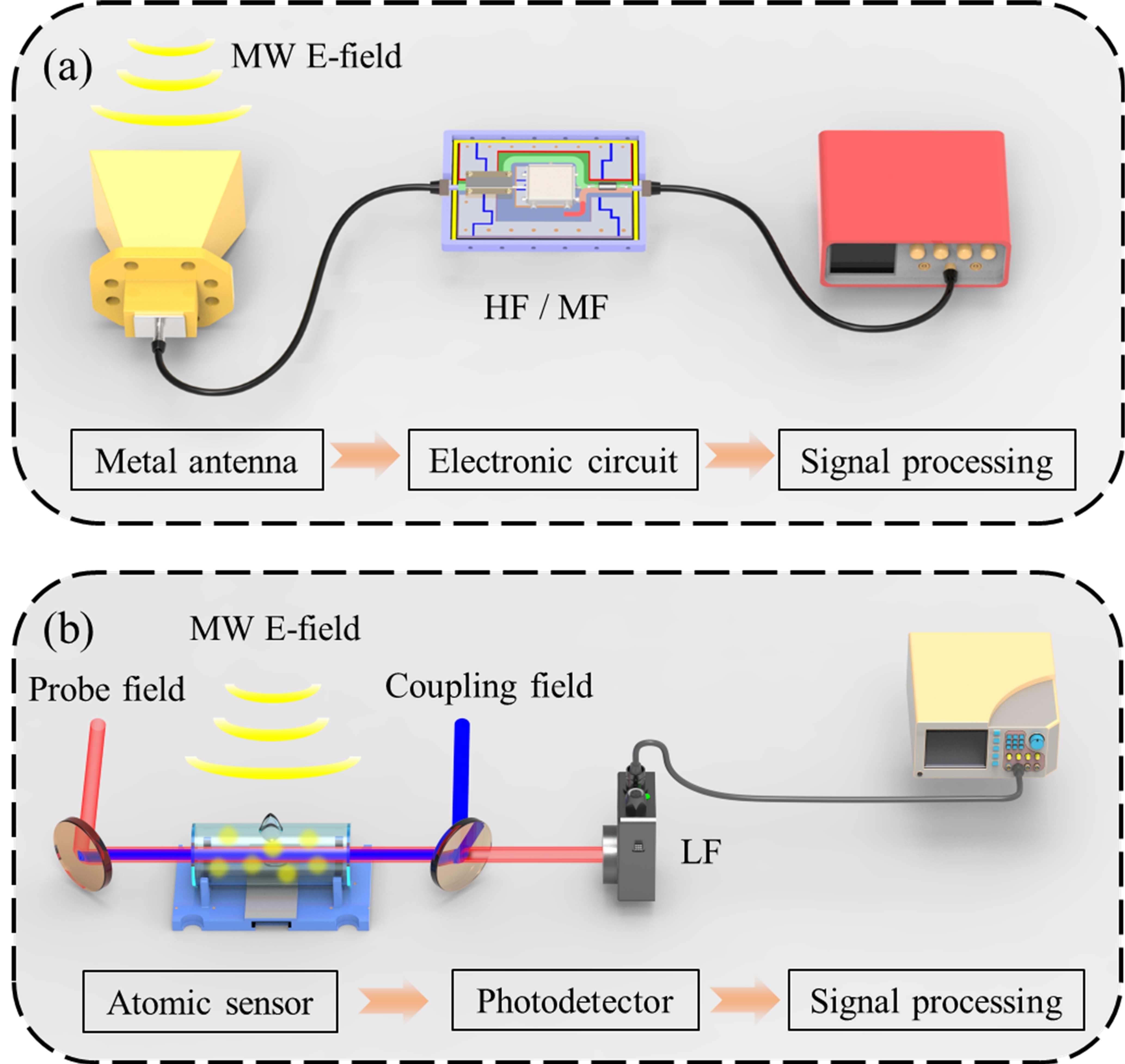}
	\centering
	\caption{The schematic diagram of (a) a traditional antenna-based MW E-field sensor and (b) an atom-based MW E-field sensor. In (a), MW E-fields are picked up by an antenna. The generated electric currents are analyzed as analog signals. In (b), lasers (probe and coupling fields) couple atomic levels (e.g., Rydberg states). The presence of MW E-fields affects laser-atom coupling. This can be used to sense weak MW E-fields using quantum coherent effects. HF, high frequency; MF, medium frequency; LF, low frequency.}
	\label{fig1}
\end{figure}

Historically, electric dipole antennas have been extensively used in MW technologies. They do not only give MW E-field applications the wings to take off over the past hundred years, but also provide means of calibration and sensing, which still occupies a dominant role in industry and life currently \cite{Kraus1988}. Metal antennas and traditional receivers for electronic circuits sit at the heart of traditional MW E-field measurement systems, as presented in \fref{fig1}(a). Microwave radiation in free space induces the vibration of  free electrons in metal antenna and are transformed into observable macroscopic currents due to electromagnetic dipole interactions. The electrical current enters electronic devices, and then becomes an analog signal before the sequence of signal processing \cite{Minasian2006}. 

This traditional approach of MW E-fields sensing has several limitations. 1) Low sensitivity. The minimum E-field strength that can be detected is about 1 mV$\cdot$cm$^{-1}$, while underlying uncertainties are large \cite{Kanda1994}. The more sensitive dipole antennas obtained through resonance enhancement come at the cost of traceability and additional calibration \cite{Basu2018}. The sensitivity of traditional metal dipole antennas is mainly affected by Johnson–Nyquist noise (also known as thermal noise), which is a current noise due to random thermal motion of charge carriers (usually electrons) inside an electrical conductor in thermal equilibrium. This type of noise is always present in the measurement environment and its intensity depends only on the temperature of electrical conductor. In the measurement, free electrons serve as signal carrier, and both signal and noise are encoded in the macroscopic motion of electrons (current). As a result, thermal noise leads to the deterioration of signal-to-noise ratio (SNR). 2) Narrow bandwidth. The traditional antennas, depending on their geometries, can only work within a finite bandwidth of frequencies, and different operating frequencies require different hardware replacements \cite{Guo2022}. 3) Complex calibration process. In order to calibrate an E-field probe, it is necessary to place the probe in an E-field with a known magnitude, but a calibrated probe is needed to identify the magnitude of this known E-field. This kind of chicken-and-egg logic makes the measurement process cumbersome. Also, it is still challenging to reliably calibrate antennas that operate at frequency above 100 GHz \cite{ArtusioGlimpse2022}. 4) Weak anti-interference ability. The material nature of bulk metal brings inevitable energy loss and poor environmental stealth. Furthermore, sizes of antennae are strictly constrained by the Chu–Harrington limit, which makes it difficult to work in restricted spaces \cite{Geyi2003}.

Quantum sensing, usually means the measurement of a physical quantity using a quantum system, quantum properties, or quantum phenomena accompanied by high sensitivity and precision, has become a growing branch of research within the area of quantum science and technology. C. L. Degen $et$ $al.$ proposed that using a quantum system with one of the following three characteristics to measure physical quantities is referred to quantum sensing \cite{Degen2017}: (I) Use of a quantum object to measure a physical quantity (classical or quantum). The quantum object is characterized by quantized energy levels. (II) Use of quantum coherence (i.e., wavelike spatial or temporal super- position states) to measure a physical quantity. (III) Use of quantum entanglement to improve the sensitivity or precision of a measurement, beyond what is possible classically. 

Recently, atom-based MW E-field sensors (shown in \fref{fig1}(b)), as an emerging technology, has the potential to overtake the antenna-based one. In the atomic sensor, a glass cell filled with alkali atomic vapor acts as a highly sensitive electric dipoles for incident microwave radiation. Uniquely the atoms are excited to the highly excited Rydberg states with a large principal quantum number through two photon transitions \cite{Gallagher1994}. The incident microwave radiation affects the internal state of Rydberg atoms, which in turn traceable superimposed on the probe laser field passing through the atom and recorded by a photodetector. Note that, the interaction between microwave radiation and Rydberg atoms is a coherent process, which is different from the absorption of incoming radiation in traditional antenna sensors \cite{Fancher2021}.

The comparison between the atom-based MW E-field sensors and the antenna-based sensors shows that the former possesses following advantages: 

1) High sensitivity. The sensitivity of using large transition dipole moments and long coherence time of Rydberg atoms make the Rydberg atom-based MW E-field sensors have higher sensitivity than traditional dipole antennas. There are typically no free electrons in the atomic system, such that the atom-based MW sensing scheme is not affected by Johnson–Nyquist noise. The decisive factor for atomic MW E-field sensors is the resolvability of spectral features with the incident MW E-field. It is fundamentally limited by the shot noise limit, whose order of magnitude is smaller than thermal noise \cite{Sedlacek2012,Fan2014}.

2) Broad bandwidth. A wide range of frequencies is covered by the plentiful energy levels of Rydberg atoms. Single species of alkali atoms have discrete to quasi-continuous energy levels that enable sensing bandwidth from megahertz to terahertz \cite{Simons2016a,Cui2023}. This capacity can be further enhanced by using multi-element collaboration. It means that one can achieve ultra-wideband measurements using only one atomic receiver hardware, which is an extremely superior feature in certain restricted spaces (satellite or spaceship, for example) \cite{Holloway2014a,Simons2016}.

3) Traceability. The very core of all electromagnetic measurements is having accurate calibrated probes and antennas \cite{2021}. Measurement results of atom-based MW E-field sensors directly link to the International System of Units (SI).The Rydberg atom-based MW E-field sensors do not need to be pre-calibrated for the unbroken traceability chain of E-field strength linking to Planck constant \cite{Holloway2014a}. The atomic transition frequency serves as a frequency reference for the Rydberg atom-based MW E-field sensor, providing the ability of frequency self-calibration. This is a promising solution to the difficulty of lacking absolute frequency standards above 100 GHz using the classic antenna method \cite{ArtusioGlimpse2022,Holloway2014a}.

4) Stealth. The glass cell, as opposed to bulk metal, has the less scattering and distortion for the microwave radiation and is advantageous in stealth occasions. On the other hand, the atomic E-field sensors can also protect itself from intense electric environment for the glass shell immune to electromagnetic fields and absorption-less mechanism. The sensor with glass cell can be connected to a shielded box that houses the lasers and other requisite electronics via optical fiber. The vapor cell is currently further optimized to its’ geometry and materials \cite{Fan2016,Fan2015a,Zhang2018}. Notably we deal with coherent coupling instead of absorbing, and optically reading changes in laser field rather than electronic readout of the current via absorbing microwave electromagnetic fields. All these mechanisms protect the sensor's performance in strong fields \cite{Fan2015a}.

5) Integration.The all-optical readout mechanism eliminates the need for ground processing and avoids the energy reflection effect due to the impedance matching. The multiple bandwidth requirements can be met by integrating multiple components into a single device, since the operating frequency is determined by desired Rydberg states that can be excited by lasers with corresponding frequencies. At the same time, the compactness and portability of atomic sensors are gradually being matured with the fiber-coupled integrated atomic vapors and the low-cost and portable lasers \cite{Holloway2017}.

In this review, we will introduce the basic mechanism of quantum sensing of MW E-field with Rydberg atoms, and the progress of this field from the perspective of progressive development of sensor. The content is organized as follows. \Sref{s2} presents the basic principle of sensing MW E-fields using the Rydberg atom. \Sref{s3} presents the progress of Rydberg atoms based MW E-field sensor in sensitivity and bandwidth. \Sref{s4} highlights the superheterodyne quantum sensing with MW-dressed Rydberg atoms. \Sref{s5} focus on the efforts for the improvement in sensitivity and bandwidth with Rydberg atoms based MW E-field sensing system.  Finally, a brief outlook is presented in \sref{soutlook}.

\section{Principle of MW E-field sensing with Rydberg atoms}
\label{s2}

Quantum sensing includes three processes, namely preparation of an initial quantum state, evolution of quantum state when interacting with physical quantities to be measured, and readout of the final state \cite{Degen2017,Tanasittikosol2011,Adams2019}. First, a controllable quantum system is prepared to a preferred initial quantum state with known properties, which usually show as the strong coupling with a physical quantity that we want to measure. Then during the evolution, the physical quantity interacts with the quantum state; hence, a final state is established. The readout of final state and comparison of the change with respect to the initial state generate information are the physical quantities to be measured. It is worth mentioning that in atomic E-field sensor, all-optical approach is employed both in the stages of preparation of Rydberg states and the readout of final state. In the three steps, the initial state preparation and the final state readout are particularly important in the atomic E-field sensor, which will be introduced in the following.

\subsection{Preparation of initial quantum states of Rydberg atoms}

Rydberg atoms are electronically excited atoms with one or more valence electrons that have a very high principal quantum number, $n$. As might be expected, such extreme atomic states possess very unusual physical and chemical properties. Since 1888 Swedish spectroscopist J. R. Rydberg first characterized properties of Rydberg atoms with the formula inscribed by his name \cite{Gallagher1994}, the investigation of these exotic atomic states has continued more than a hundred years. Recently the study of Rydberg states has expanded rapidly to new research directions.

Properties of Rydberg atoms intrinsically scale with the principal quantum number $n$ rapidly. Some of the relevant ones are summarized as follows. Their radius and dipole moment scale as $n^{2}$, while their radiative lifetime and polarizability scale as $n^{3}$ and $n^{7}$, respectively. Together with the close separation between Rydberg levels (scaling as $n^{-3}$), these lead to large two-body interactions between Rydberg atoms, scaling as $n^{4}$ and $n^{11}$ for dipole-dipole and van der Waals types, respectively \cite{Walker2012}.

As an example, we list some properties of Rb and Cs atoms in \tref{tab1}, which are most commonly used in MW quantum sensing. The numerical values of Rydberg states are huge and vastly different from ground state atoms. It is these peculiar properties of Rydberg atoms that enable the quantum sensing of MW E-field.

\begin{table}[ht!]
	\caption{\label{tab1}Rydberg state parameters used for MW E-field sensing.}
	\resizebox{\columnwidth}{!}{
		\begin{tabular}{ccccccl}
			Alkali & Rydberg                & Orbital         & Radiative         & Dipole          & Polarizability   &  \\
			atom   & state                  & radius ($\mu$m) & lifetime ($\mu$s) & moment (ea$_0$) & (MHz cm$^2$$\cdot$V$^{-2}$) &  \\ \cline{1-6}
			Rb     & 53$D_{5/2}$\cite{Sedlacek2012,Sedlacek2013} & 0.318           & 150.2             & 3623            & 217              &  \\ \cline{2-6}
			& 100$D_{5/2}$\cite{Holloway2014a}        & 1.13            & 1039.6            & 13199           & 20900            &  \\ \cline{1-6}
			Cs     & 30$D_{5/2}$\cite{Anderson2020}         & 0.065           & 13.9              & 1125            & 18.8             &  \\ \cline{2-6}
			& 52$D_{5/2}$\cite{Kumar2017a}         & 0.195           & 80.1              & 3645            & 1330             & 
		\end{tabular}%
	}
\end{table}

1) Exaggerated sensitivity to E-fields. The radius of atom is determined by the volume occupied by the electrons, which are bound to the nucleus by electrostatic forces. If an electron in the atom is given additional energy, it can jump into one of a series of allowed “excited states", whose energies and orbital motions are defined by $n$ \cite{Sevincli2014}. As the energy transfers and $n$ increases, the electron is able to travel farther from the nucleus with the distance scaling as $n^{2}$. As $n$ increases, the extra energy required to remove the electron from the atom, termed as binding energy, decreases rapidly, scaling as $n^{-2}$. As a result, high-$n$ Rydberg atoms are not only very large but are also very fragile.

\begin{figure}[ht!]
	\includegraphics[width=12.0cm]{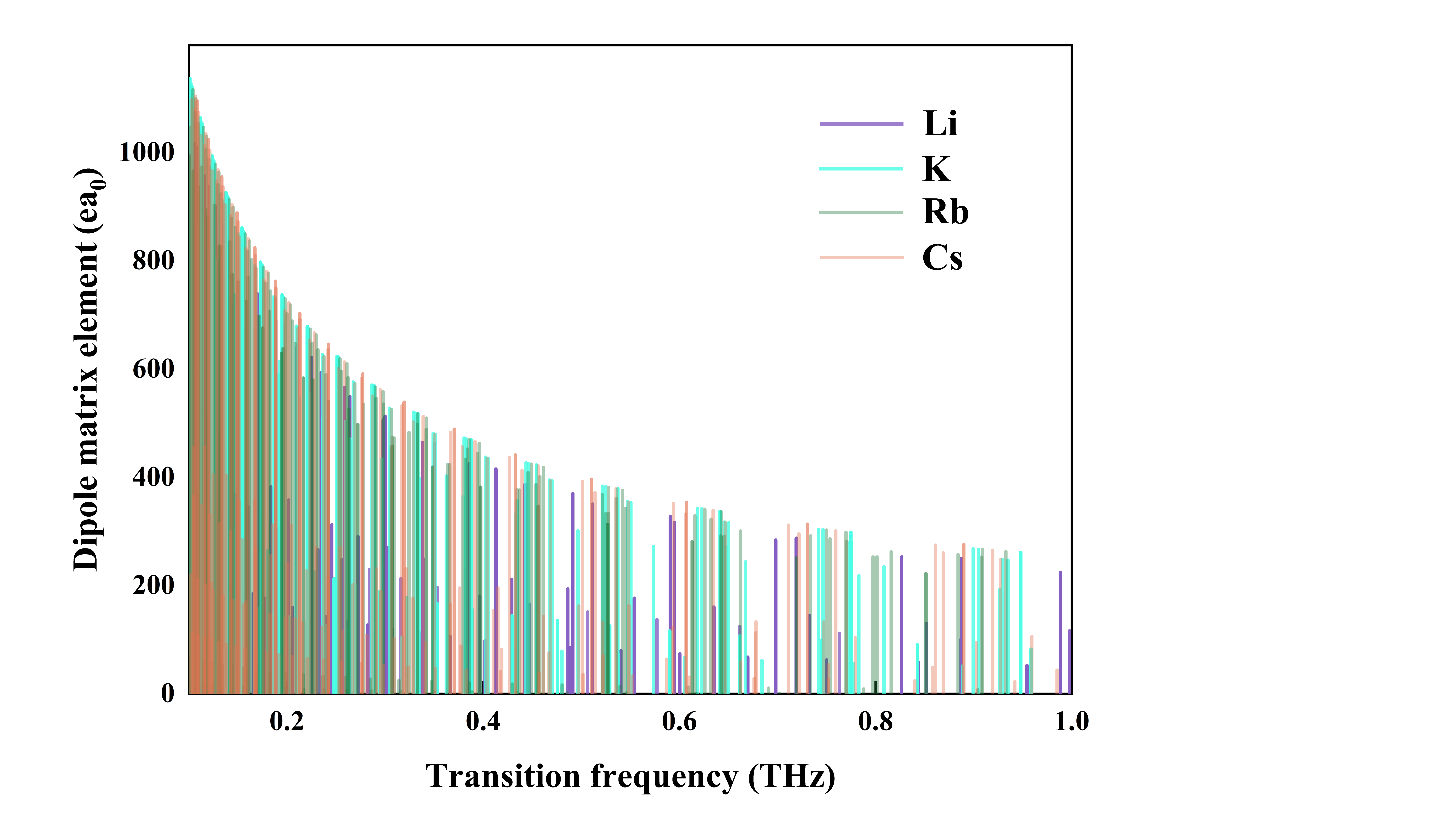}
	\centering
	\caption{The quasi-continuous transition frequencies and corresponding dipole moments from different alkali atoms covering the microwave and terahertz range. The exact transitions in Fig. 2 including n S$_{1/2}$ – (n-1)/n/(n+1) P$_{1/2, 3/2}$, n P$_{3/2}$ – n/(n-1) D$_{3/2, 5/2}$, n D$_{5/2}$ – (n-2)/n/(n+1) F$_{5/2, 7/2}$ transitions of lithium, n S$_{1/2}$ – (n-1)/n P$_{1/2,3/2}$, n P$_{3/2}$ – (n-1)/n D$_{3/2, 5/2}$, n D$_{5/2}$ – (n-2)/(n-1)/n F$_{5/2, 7/2}$, n F$_{7/2}$ - n G$_{9/2}$ transitions of potassium, n S$_{1/2}$ – n/(n+1) P$_{1/2, 3/2}$, n P$_{3/2}$ – (n-1)/n D$_{3/2, 5/2}$, n D$_{5/2}$ – (n-2)/(n-1)/n F$_{5/2, 7/2}$, n F$_{7/2}$ - n G$_{9/2}$, n G$_{9/2}$ - n H$_{11/2}$ transitions of rubidium and n S$_{1/2}$ – n/(n+1) P$_{1/2, 3/2}$, n P$_{3/2}$ – (n-1)/n D$_{3/2, 5/2}$, n D$_{5/2}$ – (n-2)/(n-1)/n F$_{5/2, 7/2}$, n F$_{7/2}$ - n G$_{9/2}$ transitions of cesium. Here n ranges from 24 to 199. }
	\label{fig2}
\end{figure}

In high-$n$ Rydberg states, the excited electron is so far away from the rest of atom (the “atomic core") that the electrostatic attraction it experiences from the nucleus is very weak, and its motion can be strongly perturbed or even dominated by weak external E-fields. In a classical picture, the loosely bound electron in a highly excited orbit is easily displaced by E-fields. Quantum mechanically, its motional states are coupled by strong electric dipole transitions and experience strong Stark shifts.

2) Plentiful energy levels. The transition frequencies of Rydberg atoms with different $n$ are expanded from megahertz to terahertz range. \Fref{fig2} shows the frequency coverage of different alkali Rydberg atoms and corresponding dipole moments. As can be seen from the figure, we can achieve the coverage of target frequency by selecting different atomic species and principal quantum numbers. Also, the small energy interval between two adjacent energy levels, scaling as $ {\Delta} E {\propto}  [1/n^{2} - 1 /(n+1)^{2}] {\approx} 2/n^{3}$, enables quasi-continuous-frequency sampling in measurements.

\begin{figure}[ht!]
	\includegraphics[width=13.5cm]{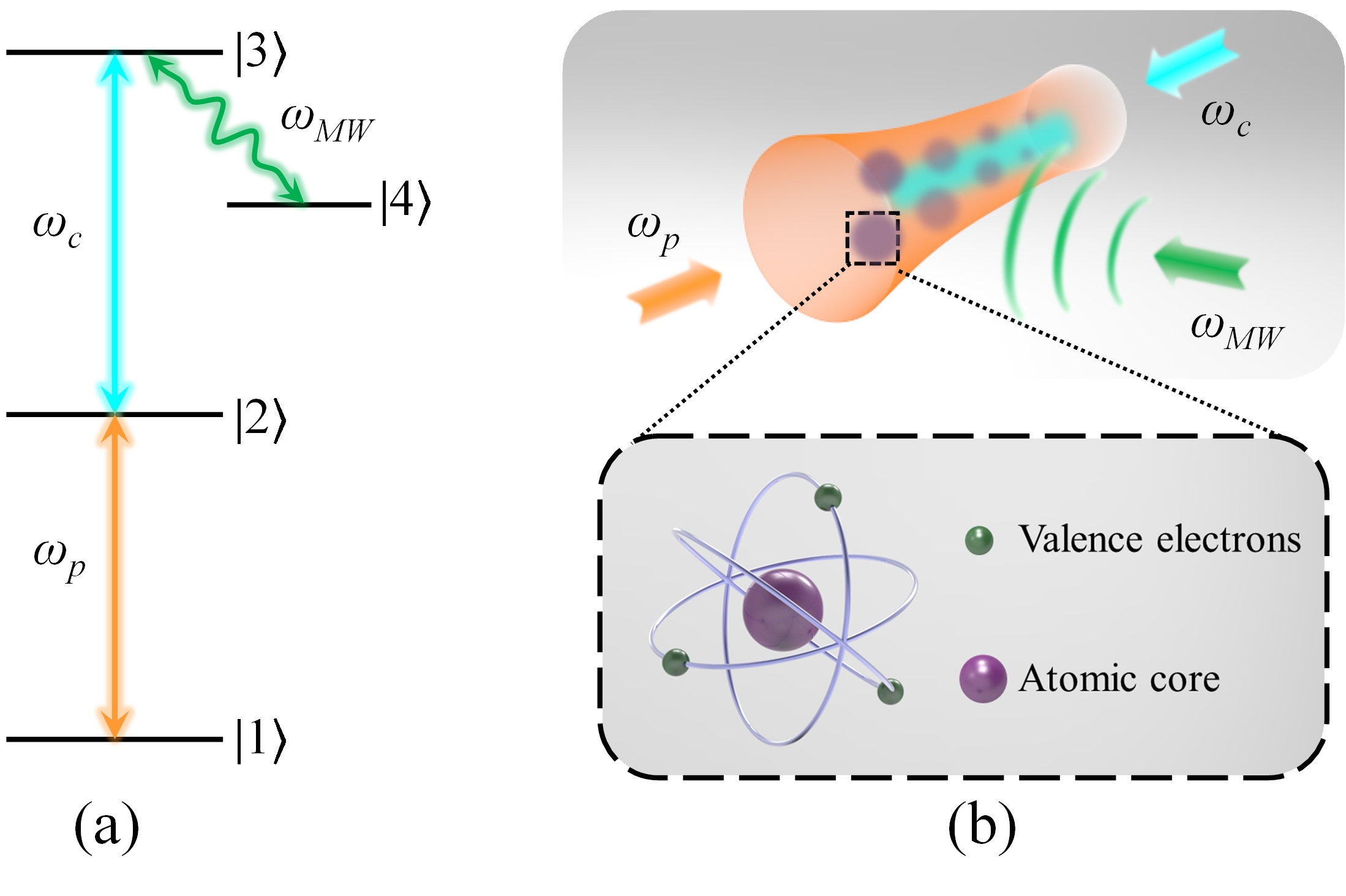}
	\centering
	\caption{(a) Typical four-level system diagram employed in MW E-field sensing. Two underlying states $\left| 1 \right\rangle$  and $\left| 2 \right\rangle$ are coupled by a probe laser field with frequency $\omega_p$. State $\left| 2 \right\rangle$  is driven by a coupling field (frequency $\omega_c$) to Rydberg state $\left| 3 \right\rangle$  coherently. A MW field with frequency $\omega_{MW}$ couples state $\left| 3 \right\rangle$ to another Rydberg state $\left| 4 \right\rangle$. (b) Schematic diagram of the Rydberg atomic preparation experimental setup. The probe and control fields counter-propagate in vapor of atoms, confined in a glass cell. The MW field affects the $\left| 1 \right\rangle \rightarrow \left| 2 \right\rangle$  transition ultimately, which can be measured from recording intensity changes of the probe laser.}
	\label{fig3}
\end{figure}

3) Mature all-optical preparation setup. The preparation of Rydberg atoms is accomplished by applying two laser beams to an atom ensemble (seen in \fref{fig3}). At the center of sensor, a vacuum glass cell consisting of a single species of alkali atoms works at room temperature, also referred as "warm atoms". Highly portable atomic cell with volume less than 1 cm$^3$ also has important applications in magnetometers, spectroscopic reference cells, and atomic frequency references \cite{Knappe2006,Petremand2012,Shi2022}. The atomic vapor cells and two-photon excitation lasers have the commercial products, which ensure the all-optical preparation of Rydberg atoms.

A typical scheme for two-photon excitation of Rydberg atoms is shown in \fref{fig3}, where two laser fields propagate in opposite directions in an atomic vapor cell. The probe laser field is used to excite atoms from the ground state ($\left| 1 \right\rangle$) to an intermediate state ($\left| 2 \right\rangle$), while the coupling laser field excites atoms from the intermediate state to a selected Rydberg state ($\left| 3 \right\rangle$). Note that, there are also alternative single-photon and three-photon excitation schemes that can be used to excite the atoms to Rydberg states \cite{Carr2012,Bai2020}.

\subsection{The readout of the final state}

The readout of final state of Rydberg atoms is usually achieved in two ways. One method is to directly measure numbers of Rydberg atoms through the ionization method. Specifically, a state-dependent E-field is applied to ionize Rydberg atoms into Rydberg ions. The ions are accelerated and detected at a multichannel plate detector. Information on the number of different Rydberg states can be obtained through controlling the ionization field strength \cite{Li2018,Lee2016}. Meanwhile, in “warm atoms”, highly excited atoms would be ionized mainly by collisions with other atoms. The ionization products are detected as a current flow between the two electrodes plates to measure the population of Rydberg states \cite{Barredo2013}. The ionization destroys quantum states and coherence of atoms, which is considered as a disposable resource and is therefore destructive measurements. In the meantime, the whole setup is complex, making it difficult for applications outside the laboratory. The other method, based on quantum interference (e.g., electromagnetically induced transparency (EIT) and electromagnetically induced absorption (EIA) effects), is an all-optical non-destructive readout methods. The information of final state is recorded in the amplitude and phase of probe laser after interacting with Rydberg atoms. The all-optical method only drives transition between atomic internal-states without destroying the atoms (i.e., ionization), which can be recycled in continuous measurement \cite{Johnson2010,Li2019}.

\begin{figure}[hb!]
	\includegraphics[width=15.0cm]{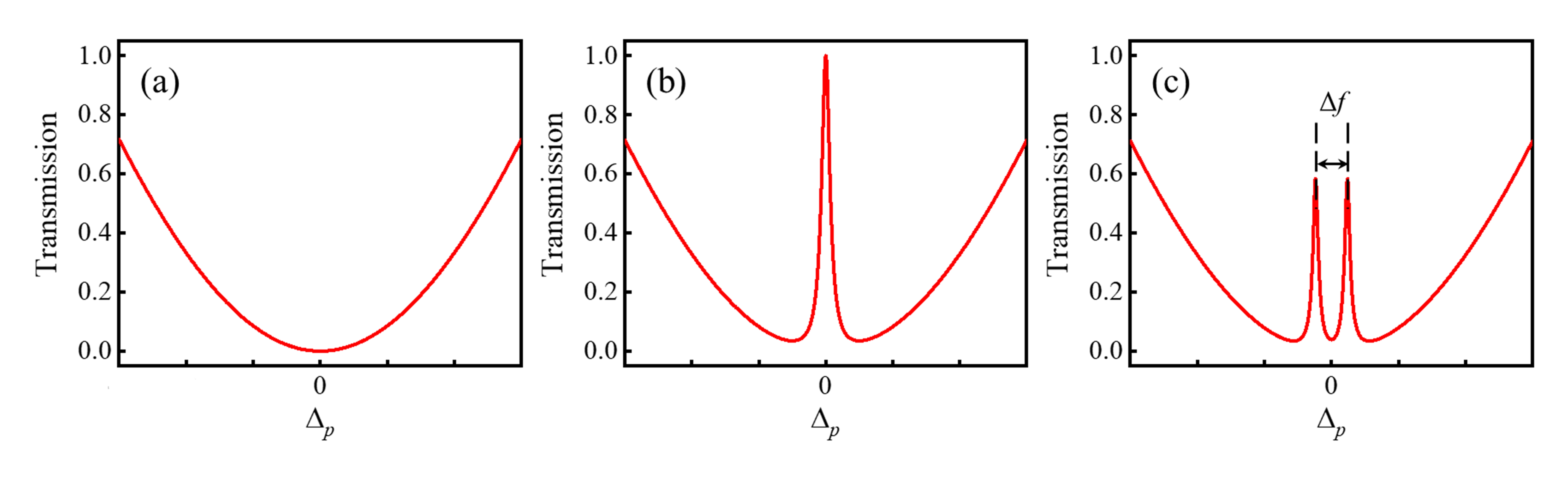}
	\centering
	\caption{Probe laser spectrum evolution affected by incident fields. (a) Absorption occurs when the resonant probe laser passes through the atoms. (b) Electromagnetically induced transparency (EIT) associated with the probe and coupling lasers in a three-level system. (c) Autler-Townes (AT) splitting with a MW E-field corresponding to a transition between two Rydberg states introduced.}
	\label{fig4}
\end{figure}

Coupled by an incident MW field, atoms in Rydberg state ($\left| 3 \right\rangle$) are on or near resonance with the transition to a nearby Rydberg state ($\left| 4 \right\rangle$). Thus, properties of the incident microwave field can be measured via optical readout of the probe beam followed by spectral analysis. \Fref{fig4} shows effects of the coupling and MW field on the transmission spectrum of probe beam. Without other fields, the probe laser experiences strong absorption, shown in \fref{fig4}(a), when its frequency is tuned close to resonance. Note that the spectrum in a larger frequency sweep range is similar as reported in the ref. \cite{Mohapatra2007}. When the coupling field is turned on, a fraction of the atoms is excited to the Rydberg state. A phenomenon known as EIT is observed, where a narrow window with high transmission is opened in the spectrum of probe beam, as shown in \fref{fig4}(b). In this narrow window, we can intuitively understand that the probe beam passes through the atomic medium but is not absorbed. In a quantum picture, a dark state, refers to a state of an atom cannot absorb photons, exists in this process as the result of quantum interference in this three-level system. When an atom is in a coherent superposition of two states, both of which are coupled by lasers at the right frequency to a third state, the system can be made transparent to both lasers as the probability of absorbing a photon goes to 0. Finally, the presence of a microwave field on or near resonance with an adjacent Rydberg state induces Autler-Townes (AT) splitting. Here the alternating current (AC) Stark effect splits the EIT transmission window, as shown in \fref{fig4}(c).  The AT splitting region is the most sensitive region of MW E-field sensors because resonance enhancement is achieved. The frequency shift $\Delta f$ observed in the spectrum of probe beam is related to the strength of MW field and the dipole moment of the transition:
\begin{equation}
	\Delta f=k\cdot\Omega _{MW}/2\pi.
	\label{equ1}
\end{equation}
where the ratio $k$ accounts for the Doppler mismatch between the two fields. Ratio $k$ will be 1 when scanning the coupling field, and is $\lambda_p/\lambda_c$ when scanning the probe field, respectively. In the above equation, $\Omega_{MW}$ is the Rabi frequency of the microwave field, given by
\begin{equation}
	\Omega_{MW}=\mu_{MW}\cdot E_{MW}/\hbar.
	\label{equ2}
\end{equation}
Here, $\mu_{MW}$ is the transition dipole moment, $E_{MW}$  is the amplitude of microwave field, and $\hbar$ is reduced Planck’s constant.

Thus, we obtain a very important relation: $E_{MW}=\hbar\Delta f/\mu_{MW}$. This is the basic principle of sensing the MW E-field using the AT splitting of Rydberg-EIT spectrum. It shows that the amplitude information of MW field can be obtained by spectral analysis and is directly traceable to Planck's constant.

\section{Progress of sensitivity and bandwidth in Rydberg atoms based MW E-field sensing system}
\label{s3}
\begin{figure}[ht!]
	\includegraphics[width=13.0cm]{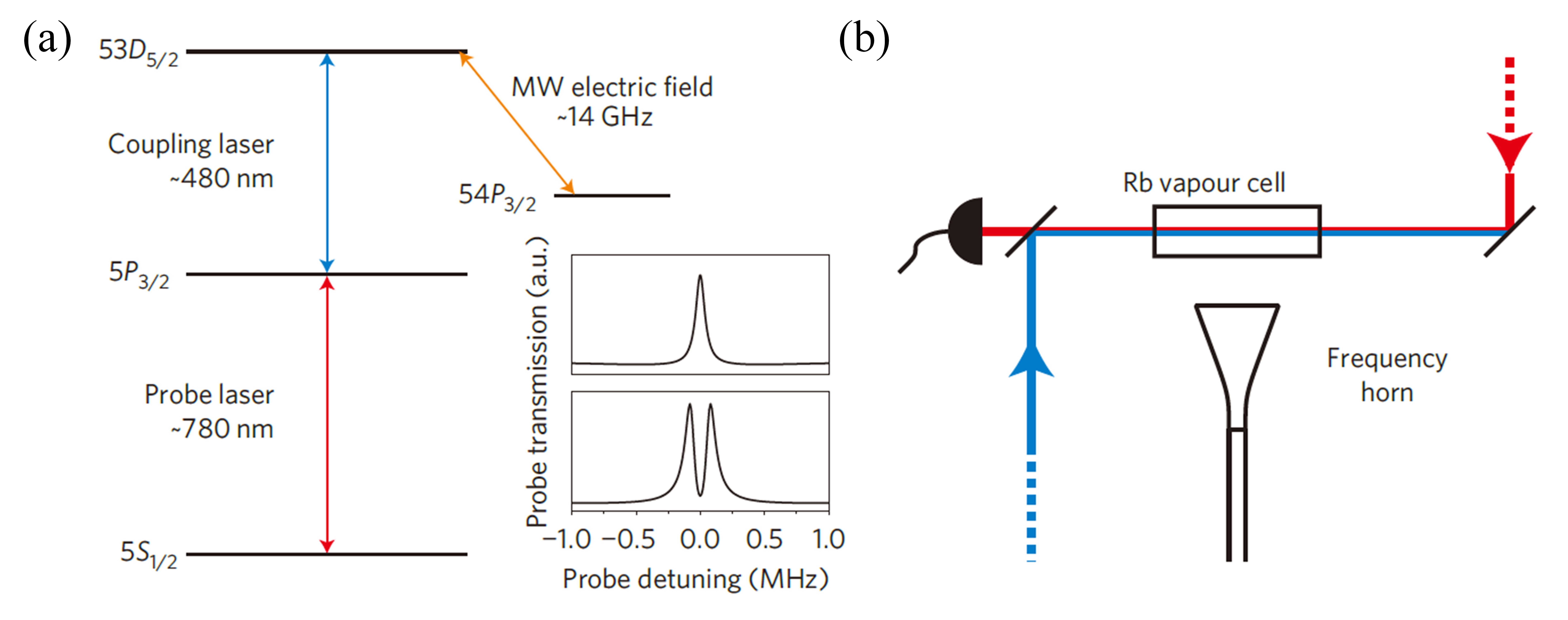}
	\centering
	\caption{(a) Energy level diagram of four-level system used for the MW E-field sensing \cite{Sedlacek2012}. The top part of inset shows an EIT spectrum in the three-level system without a MW E-field. The bottom part of inset shows an AT splitting spectrum when a MW E-field is present. (b) Diagram of experimental setup used in the experiments.}
	\label{fig5}
\end{figure}

In 2012, the researchers from the university of Oklahoma and the university of Stuttgart first demonstrated MW E-field sensing at frequency of 14.23 GHz by the AT splitting of EIT spectrum with a 5$S_{1/2}$$\rightarrow$5$P_{3/2}$$\rightarrow$53$D_{5/2}$$\rightarrow$54$P_{3/2}$ four-level $^{85}$Rb atomic coherent system \cite{Sedlacek2012}. The diagram of energy level and experimental setup of this system are shown in \fref{fig5}. In this Rydberg atoms based MW E-field sensing setting, the measurement of MW E-field strength is converted into the optical frequency discrimination, which is directly traceable to Planck’s constant. The minimum detected MW E-field is down to 8 $\mu$V$\cdot$cm$^{-1}$ with a sensitivity of 30 $\mu$V$\cdot$cm$^{-1}$Hz$^{-1/2}$, which is already superior to present records of traditional antenna-based sensors.

\begin{figure}[ht!]
	\includegraphics[width=16.0cm]{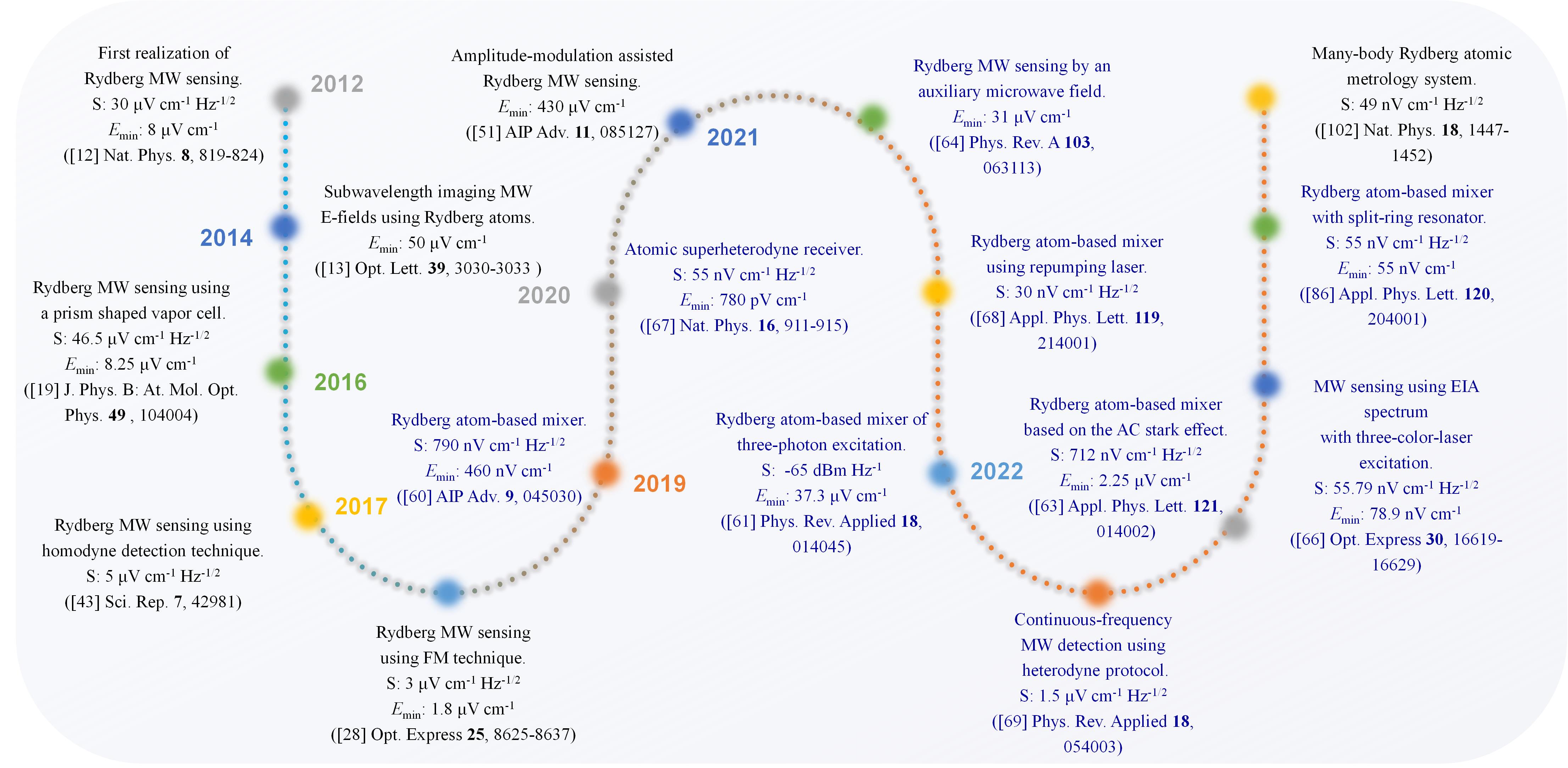}
	\centering
	\caption{The main progress of Rydberg atoms based MW E-field sensing in sensitivity. S: sensitivity; $E_{min}$: The minimum detectable strength of MW E-field. The black texts represent Rydberg atoms based MW E-field sensing in the classical four-level system, the blue texts represent Rydberg atoms based MW E-field sensing with an auxiliary MW E-field introduced.}
	\label{fig6}
\end{figure}

Among core indicators of Rydberg atoms MW E-field sensing technologies, sensitivity and bandwidth have attracted significant attention at first. Better performance in these indicators can be achieved in comparison with traditional electrometry. The sensitivity of Rydberg atoms based MW E-field sensing is mainly limited by the technical noises of system and the spectrum resolution of AT splitting. The minimum detectable electric field in the AT splitting region is given by \cite{Fan2015}:

\begin{equation}
	E_{min} =\frac{h}{\mu_{MW}T_{meas}\sqrt{N}}.
\end{equation}
where $h$ is Planck's constant, $T_{meas}$ is the measurement time, and $N$ is the number of independent measurements. Thus, the minimum detectable field is proportional to the minimum resolvable phase shift and inversely proportional to the product of interaction strength and the square root of measurement time. Considering practical factors and guided by theoretical analysis, improvement of the sensing sensitivity is achieved through two routes, namely increasing the spectral SNR through suppressing the technical noises of system, and improving the AT splitting spectral resolution by exploring new mechanisms \cite{Fan2015}. \Fref{fig6} gives the main progress in the measurement sensitivity (black text).

The noise on the optical fields and in the signal readout path (photodetector and electronics) always exists when the spectra are measured by optical fields. Therefore, it becomes highly desirable to suppress or even eliminate these noises in quantum sensing of MW E-fields with Rydberg atoms. Optimizing the geometric structure of sensing system is first brought to mind. It is found that atomic vapors with a special geometric structure can reduce noise to a certain extent, which can be applied in spatial distribution measurement of MW E-field with high resolution \cite{Fan2014,Holloway2014}. The detailed study indicates that a cell size far smaller than the wavelength of MW E-field can eliminate the Fabry–Pérot effect of vapor cell and improve the measurement accuracy \cite{Fan2015a}. This result is also well verified in two-dimensional high-resolution spatial distribution of MW E-field strength inside a cubic cell and a cylindrical cell \cite{Zhang2018}. On this basis, a prism-shaped atomic vapor was introduced into the atomic MW E-field sensing system. It utilizes the dispersive part of EIT effect. The MW E-field induces changes to the refraction index of vapor resulting in deflection of the probe laser beam. The minimum detectable strength of MW E-field is 8.25 $\mu$V$\cdot$cm$^{-1}$ with a sensitivity of 46.5 $\mu$V$\cdot$cm$^{-1}$Hz$^{-1/2}$, which is comparable to the absorptive part of EIT readout signal \cite{Fan2016}. A free-space Mach-Zehnder interferometer along with a homodyne detection technique, which detects the non-linear phase shift instead of directly measuring the transmitted probe power, is put forward to improve the sensitivity. In this way, the noise of probe laser is reduced by the subtraction taking place in the homodyne detection and the EIT signal is enhanced by the strong local field, and finally achieved a sensitivity of 5 $\mu$V$\cdot$cm$^{-1}$Hz$^{-1/2}$ \cite{Kumar2017}. Moreover, comprehensive experimental and theoretical studies on three-photon and four-photon Rydberg-EIA/EIT in an atomic vapor cell provides us a better understanding of the sensing mechanism and approaches for optimization \cite{Kondo2015,Thaicharoen2019}. Frequency modulation spectroscopy with active control of residual amplitude modulation, which can eliminate technical noise in the probe laser readout and elucidate the role of probe laser shot noise, was then applied to MW E-field sensing. Finally, a minimum detectable strength of 1.8 $\mu$V$\cdot$cm$^{-1}$ with sensitivity of 3 $\mu$V$\cdot$cm$^{-1}$Hz$^{-1/2}$ is realized \cite{Kumar2017a}. The joint use of frequency modulation, amplitude modulation and homodyne detection methods promotes further developments. The $1/f$ noise of spectrum is significantly suppressed and spectral SNR is improved by 5.8 times, and a factor of 2 measurement sensitivity improvement is achieved \cite{Li2020}. A hybrid atomic sensor with a passive resonator element embedded in the vapor cell provides a 24 dB gain enhancement in intensity sensitivity and polarization selectivity for directional field detection \cite{Anderson2018}. The optimal angular-state choice for the measurement of MW E-fields with different strengths is also a boost to the sensitivity improvement \cite{Chopinaud2021}. 

The sensitivity of Rydberg atoms based MW E-field sensing system also depends on spectral resolution of the AT splitting. For weak MW fields, it is difficult to detect and measure the AT splitting in the EIT signal. This problem becomes especially relevant in the sensing of high-frequency MW E-fields because high frequency measurement requires low Rydberg states, whose dipole moments are significantly smaller than those of higher $n$ states used in sensing low frequency MW fields. A frequency detuning method is introduced to mitigate this issue, which is accomplished by varying the MW frequency around a resonant atomic transition and extrapolating the weak on-resonant field strength from the resulting off-resonant AT splitting \cite{Simons2016, Zhang2019}. As a result, the improvement of greater than a factor of 2 in the measurement sensitivity compared with on-resonant AT splitting MW E-field measurement is achieved \cite{Simons2016}. The Zeeman frequency modulation spectroscopy conducted by an AC modulated magnetic field is demonstrated to have a SNR enhancement of more than 10 times compared with the original spectrum, resulting the detection of MW E-fields as small as 33 $\mu$V$\cdot$cm$^{-1}$ \cite{Jia2020}. The amplitude modulation of MW field achieves six-time improvement in EIT-AT splitting resolution, where the minimum detectable MW E-field strength is 430 $\mu$V$\cdot$cm$^{-1}$ \cite{Liu2021a}. The three-photon readout scheme, which reduces the residual Doppler shifts to the order of magnitude of the Rydberg state decay times, extends the AT splitting regime of Rydberg atoms based radio frequency (RF) electrometry to sensing lower RF E-field strengths \cite{Shaffer2018,Ripka2021}. However, the optimum sensitivity of quantum sensors based on EIT of three-level system was derived, showing clear boundaries. It reveals that ladder-EIT system cannot achieve the standard quantum limit due to unavoidable absorption loss. Such predication calls for new approaches and mechanisms to overcome the limit \cite{Meyer2021}.

\begin{figure}[ht!]
	\includegraphics[width=15.0cm]{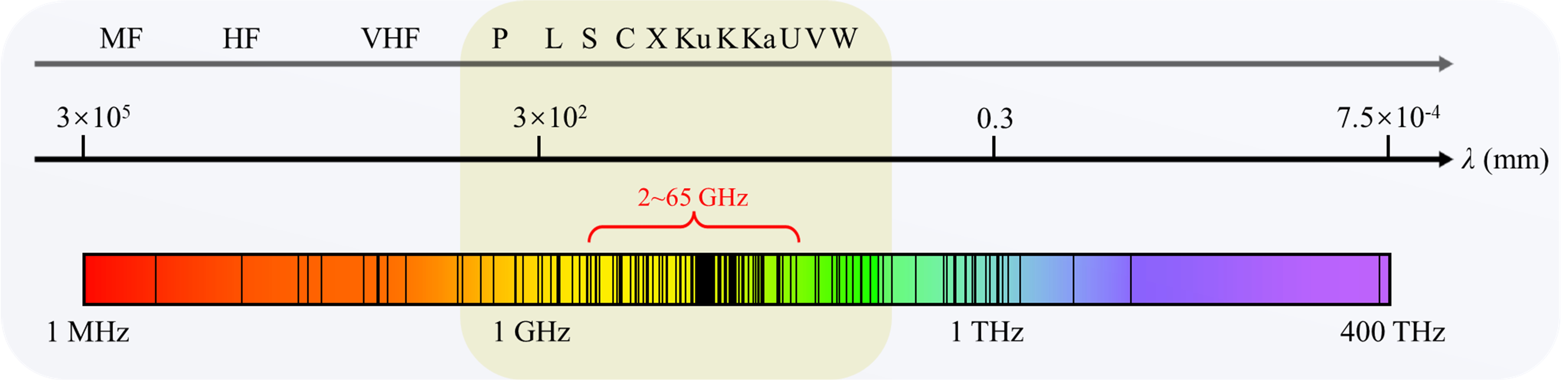}
	\centering
	\caption{The reported operating frequencies of Rydberg atoms based MW E-field sensing system. Black lines represent for valid sensing results from Rydberg atoms based electrometer. Sensing results cover most of frequency range, and mainly focus on a practical frequency band from 2 to 65 GHz.}
	\label{fig7}
\end{figure}

In the ongoing pursuit of higher sensitivity of MW E-fields, the operating bandwidth, is another important performance indicator to be investigated. The plentiful energy levels of Rydberg atoms cover wide frequency ranges, which renders it a perfect candidate for broadband quantum sensing over the range from direct current (DC) to THz \cite{Holloway2014a}. Especially, using Rydberg atoms based MW E-field sensing system, fields calibration in frequencies above 100 GHz are established, where there is no traceable standard currently \cite{Gordon2014, Tonouchi2007, Lam2021}. \Fref{fig7} gives an overview of the experimentally reported frequency coverage, which shows that although the system can work in frequency bands hopping within a wider frequency range, the operating bandwidth is still a series of discrete frequencies around the atomic transition frequencies from $\sim$GHz to $\sim$THz with a narrow bandwidth. The operating bandwidth of Rydberg atoms based MW E-field sensing can be widened by multiplexing atomic species \cite{Simons2016a} or frequency-division-multiplexing technique \cite{Zou2020}. Theoretical investigation shows that real-time detection of the absolute strength of terahertz E-fields with 0-5 THz bandwidth can be achieved at room temperature in a four-level Rydberg atomic setting~\cite{Zhou2022}.

\section{The superheterodyne quantum sensing with microwave-dressed Rydberg atoms}
\label{s4}
Although various noise suppression and spectral resolution improvement methods are proposed and implemented, the measurement sensitivity of typical four-level system with single MW E-field just approaches to the magnitude of $\mu$V$\cdot$cm$^{-1}$Hz$^{-1/2}$. Introducing another E-field resonant or detuned at two Rydberg levels provides a new perspective for improving the sensitivity of Rydberg atoms based MW E-field sensing system. Through the use of AT splitting and a local oscillator (LO) field, the frequency splitting measurement can be converted into a modulated amplitude measurement. The main progress in this regard is shown in \fref{fig6} with blue text.

The primary work in this field gives a heterodyne detection scenario with two E-fields injecting on a vapor cell. One E-field drives a transition between adjacent Rydberg states acting as a LO field, and the second field detuned by 90 kHz from the LO field acts as the signal field. The configuration makes the Rydberg atoms as an RF mixer for weak E-field detection, such that the strength down to 460 nV$\cdot$cm$^{-1}$ is detected with a sensitivity of 790 nV$\cdot$cm$^{-1}$Hz$^{-1/2}$ \cite{Gordon2019}. The usage of heterodyne technique to amplify the system response to a 30 MHz weak signal E-field accompanied by applying a LO E-field is also reported. It achieves the minimum detectable electric field strength to be 37.3 $\mu$V$\cdot$cm$^{-1}$ with a sensitivity up to -65 dBm$\cdot$Hz$^{-1}$ and a linear dynamic range over 65 dB \cite{Liu2022a}. Also, this mechanism is employed to improve the detection sensitivity of a 500 MHz far-detuned RF field with Rydberg atoms dressed by a near-resonant MW E-field \cite{Yao2022}. On this basis, an off-resonant heterodyne method, where a strong far off-resonant field acts as a LO field and the incident weak signal field with a few hundreds of kHz difference from the LO field is mixed in the Rydberg atom vapor cell. The LO field shifting the Rydberg level to a high sensitivity region so that the incident weak field generates an intermediate frequency signal, resulting to the minimum detectable field strength of 2.25 $\mu$V$\cdot$cm$^{-1}$ with sensitivity of 712 nV$\cdot$cm$^{-1}$Hz$^{-1/2}$ of the MW E-field \cite{Hu2022}. In the higher frequency range, with an auxiliary MW field that detuned from the target field by approximately 1.4 GHz introduced into a four-level Rydberg atoms based MW E-field sensing system, the performance of this quantum microwave electrometry is improved to be 31 $\mu$V$\cdot$cm$^{-1}$ for enlarged EIT-AT splitting \cite{Jia2021}. A similar setup with the introduction of a resonant microwave field but further optimization of power ratio of the two microwave fields is also investigated, achieving a minimum detectable MW E-field of 6.71 $\mu$V$\cdot$cm$^{-1}$ \cite{Yuan2022}. The three-laser Rydberg EIA based heterodyne MW sensor gives a minimum detectable E-field of 78.9 nV$\cdot$cm$^{-1}$ in time scale of 500 ms and sensitivity of 55.79 nV$\cdot$cm$^{-1}$Hz$^{-1/2}$ \cite{You2022}.

\begin{figure}[hb!]
	\includegraphics[width=16.0cm]{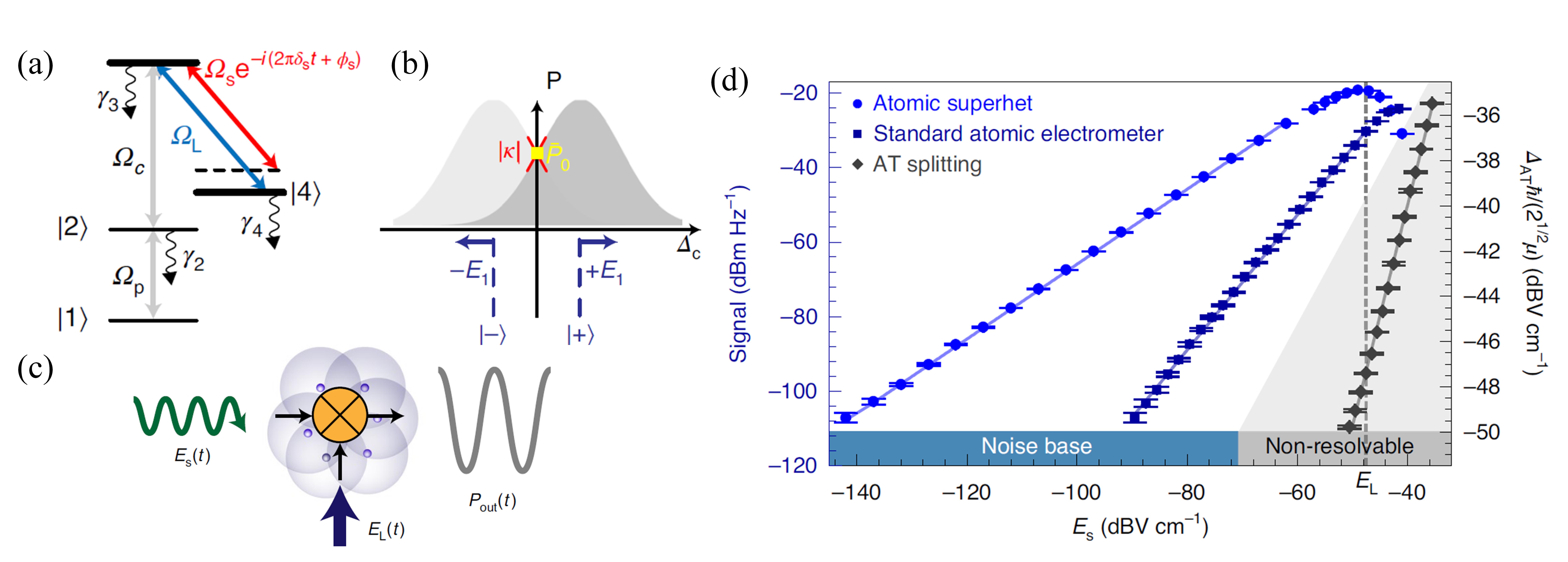}
	\centering
	\caption{  Basic principle of the Rydberg atoms based superheterodyne \cite{Jing2020}. (a) Level scheme. States $\left| 1 \right\rangle$ ,$\left| 2 \right\rangle$, and Rydberg state$\left| 3 \right\rangle$ are resonantly coupled by probe ($\Omega_p$) and control ($\Omega_c$) fields, respectively. A LO MW E-field (blue) drives Rydberg transition $\left| 3 \right\rangle \rightarrow \left| 4 \right\rangle$ with $\Omega_L$. A weak signal MW (red) yields a coupling ($\Omega_S$) related to the LO field. (b) Schematic of MW sensing with MW-dressed Rydberg EIT spectroscopy. (c) Atomic superheterodyne consisting of Rydberg atoms dressed by a LO MW $E_L(t)$ detects a MW signal $E_s(t)$ as an optical output $P_{out}(t)$. (d) Comparison of the sensitivities of atomic electric electrometer and atomic superheterodyne measurement systems.}
	\label{fig8}
\end{figure}

However, what really brought this scheme to a head is the experimental demonstration of an atomic superheterodyne receiver by research group at Shanxi University, which renders a sensitivity of Rydberg atoms based MW E-field sensing system to approach the quantum projection noise limit \cite{Jing2020}. The level scheme and basic principle of the Rydberg atoms based superheterodyne is shown in \fref{fig8}. In this scheme, a strong on-resonant field, denoted as a LO field, induces the AT splitting. A weak signal field with a few hundred kHz difference from the LO field is mixed with the LO field in the Rydberg atom vapor cell to generate an intermediate frequency signal, which is read out by Rydberg-EIT spectroscopy. The strong on-resonant LO MW field results in two dressed states $\left|\pm\right\rangle$ energetically separated by $\hbar \Omega_L$, where $\Omega_L$ is the Rabi frequency of LO field. When a weak signal MW field is applied, the energy shift $\left|\pm\right\rangle$ of the two dressed states is $\pm E_1=\pm\hbar\Omega_{S}\cos(2\pi\delta_St+\phi_S)/2$. Where $\Omega_S$ is the Rabi frequency of signal field, $\delta_s$ is the amount of frequency detuning of signal field relative to the LO field, and $\phi_S$ is the phase difference between the signal and the LO MW fields. For $\Omega_S\ll\Omega_L$, the relationship between the probe laser transmission change $P_{out}$ and the signal MW E-field is $P_{out}=\left| P(\delta_S)\right| \cos(2\pi\delta_S t+\phi_S)$. Here, $P_{out}$ is the probe lasers transmission change detected by the photodiode detector, and $P(\delta_s)$ is the amplitude of single-sided Fourier spectrum to a frequency $\delta_S$. The Rabi frequency $\Omega_S$ of signal MW filed is described as $\Omega_S=\left| P(\delta_S)\right|/k_0$. The measurement of signal MW E-field strength is converted to evaluation of the beat frequency signal strength. In this approach, the dressing field does not require external field calibration in a standard field as the dipole antenna does. While, the usage of self-calibration process by optimized LO MW E-field helps us to improve the sensitivity for signal MW E-field sensing.

The Rydberg atoms based superheterodyne receiver further improves sensitivity through reducing classical noise or accessing smaller fields by increasing the measurement time. The amplitude noise and polarization noise of the laser source are suppressed by the servo system and balanced detection technique. The transit noise due to thermal atoms is eliminated using larger-diameter probe and coupling laser beams. The laser frequency noise is eliminated by using state-of-the-art lasers with mHz line-width and expanding the servo bandwidth to several MHz. The amplifier noise of photon detector is reduced through optical heterodyne or homodyne detection. The spectrum analyzer noise is removed by utilizing conventional electronic amplifiers. After eliminating relevant technical noise, the Rydberg atoms based superheterodyne approaches quantum projection noise-limited sensitivity of $\sim$700 pV$\cdot$cm$^{-1}$Hz$^{-1/2}$. The minimum detectable electric field of 780 pV$\cdot$cm$^{-1}$ is achieved with the sensitivity of 55 nV$\cdot$cm$^{-1}$Hz$^{-1/2}$.

Further to this primary work, a ground state repumping laser beam, which increases the number of Rydberg atoms in the EIT interaction while avoiding additional Doppler-, power-, or collisional-broadening from other methods was introduced. An improvement on the sensitivity of scheme by a factor of nearly 2 and the minimum field detectable of 30 nV$\cdot$cm$^{-1}$Hz$^{-1/2}$ were achieved \cite{Prajapati2021}. A multi-level heterodyne measurement scheme, where two off-resonant MW fields acting as a tunable LO field, was proposed to give the amplitude, phase, and frequency information of the signal MW E-field. The experiment demonstrates a sensitivity of up to 1.5 $\mu$V$\cdot$cm$^{-1}$Hz$^{-1/2}$, 80 dB linear dynamic range \cite{Liu2022b}.

The atomic super-heterodyne method not only improves the measurement sensitivity, but also provides the access to frequency and phase sensing. However, instantaneous bandwidth is still main direction of efforts. The resonant sensing of MW E-fields by EIT and AT splitting in Rydberg atoms is typically limited to frequencies within the narrow bandwidth of a Rydberg transition. By applying a second field resonates with an adjacent Rydberg transition, far-detuned fields can be detected through a two-photon resonant AT splitting. This method expands the operating bandwidth from $\pm$50 MHz to the order of hundreds of megahertz by either varying the frequency or the energy level of adjacent Rydberg resonance tuning field \cite{Simons2021}.  A large progress is presented with an off-resonant RF heterodyne technique in thermal Rydberg atoms coupled to a planar microwave waveguide, resulting in a continuous operation for carrier frequencies ranging from DC to 20 GHz. The system also achieves over 80 dB of linear dynamic range \cite{Meyer2021a}. The multi-level heterodyne measurement scheme not only gives the sensitivity improvement mentioned above, but also realizes the continuous frequency range of over 1 GHz \cite{Liu2022b}. Recently, a MW-frequency-comb field is introduced into the four-level sensing system to give this old question a new answer. The novel scheme provides a real-time and absolute-frequency measurement with a range of 125 MHz \cite{Zhang2022}. 

In addition to the operating bandwidth, another mountain that needs to be climbed on the road to the general application of Rydberg atoms based MW E-fields sensing system is the instantaneous bandwidth. It is the 3 dB bandwidth for beat-note or modulated-signal response, and depends on the lifetime of atomic level \cite{Cox2018}. Recent research results show that the noise source within the instantaneous bandwidth range is mainly quantum projection noise in the Rydberg atoms based MW E-field sensing mechanism \cite{Wang2023}. The superheterodyne mechanism is used to study this parameter first, and the instantaneous bandwidth of $\pm$150 kHz was obtained \cite{Jing2020}. A 0.8 MHz instantaneous bandwidth was achieved using the AC Stark shift effect and heterodyne method for a 30 MHz MW E-field detection \cite{Liu2022a}. Combining the resonant heterodyne and two-photon AT splitting schemes improves this bandwidth into 1.9 MHz \cite{Liu2022b}. Also, the off-resonant RF heterodyne technique was used to realise 4 MHz instantaneous bandwidth \cite{Meyer2021a}. Methods of simultaneous multi-band demodulation using multiple Rydberg states is a recent effort achieving instantaneous bandwidth of 6.1 MHz \cite{Meyer2023}. Spatially splitting and staggering pulsed probe laser makes a new milestone that ensured continuous sampling of the microwave source with improved instantaneous bandwidth of 100 MHz \cite{Knarr2023}.

At the same time, this dual-MW fields geometric configuration provides additionally a research platform for measuring microwave field phases. Sensitive phase recognition relies on the relationship between the probe laser transmission change and the signal MW field. The phase variation between the signal and LO MW fields is reflected by intermediate frequency signal, which is transferred directly to an optical field. Taking the Rydberg atoms as a mixer, who demodulates the MW field and down-converts the 20 GHz MW field into an oscillation in the probe laser intensity, the phase between two MW fields is converted to an intermediate frequency on the order of kHz and readout clearly \cite{Simons2019}. Besides, the angle-of-arrival (AOA) of MW field, which is of great importance to radar and advanced communications applications, can be detected by heterodyne technique \cite{Robinson2021}. The determination of AOA is achieved by establishing the relationship between angle, position and phase for each microwave fields. These sensitive detections of MW E-field strength, phase, frequency and AOA make it possible to achieve multi-frequency and multi-dimensional detection, radar identification and microwave communications.

\section{New frontiers of Rydberg atoms based MW E-field sensing system}
\label{s5}
Along with the use of above-mentioned typical Rydberg atomic systems, EIT and AT splitting optical readouts to sense MW E-fields, some potential methods or techniques take advantage of the Rydberg mechanism but slightly different have also been proposed to realize microwave sensing with higher sensitivity and wider bandwidth. On this road, breaking through the sensitivity of photon shot noise or atomic projection noise, quasi-continuous operating bandwidth, and the instantaneous bandwidth close to the natural lifetime of Rydberg atoms are still standing at as milestones waiting for us to pursue \cite{Cox2018}. In this section, we elaborate on this topic from three aspects, the progressive incremental measurement system, the alternative Rydberg measurement medium and the quantum-enhanced measurement system.

\subsection{The progressive incremental measurement system}

\hspace*{\fill}

1)  Cavity-enhanced measurement.

\begin{figure}[ht!]
	\includegraphics[width=15.0cm]{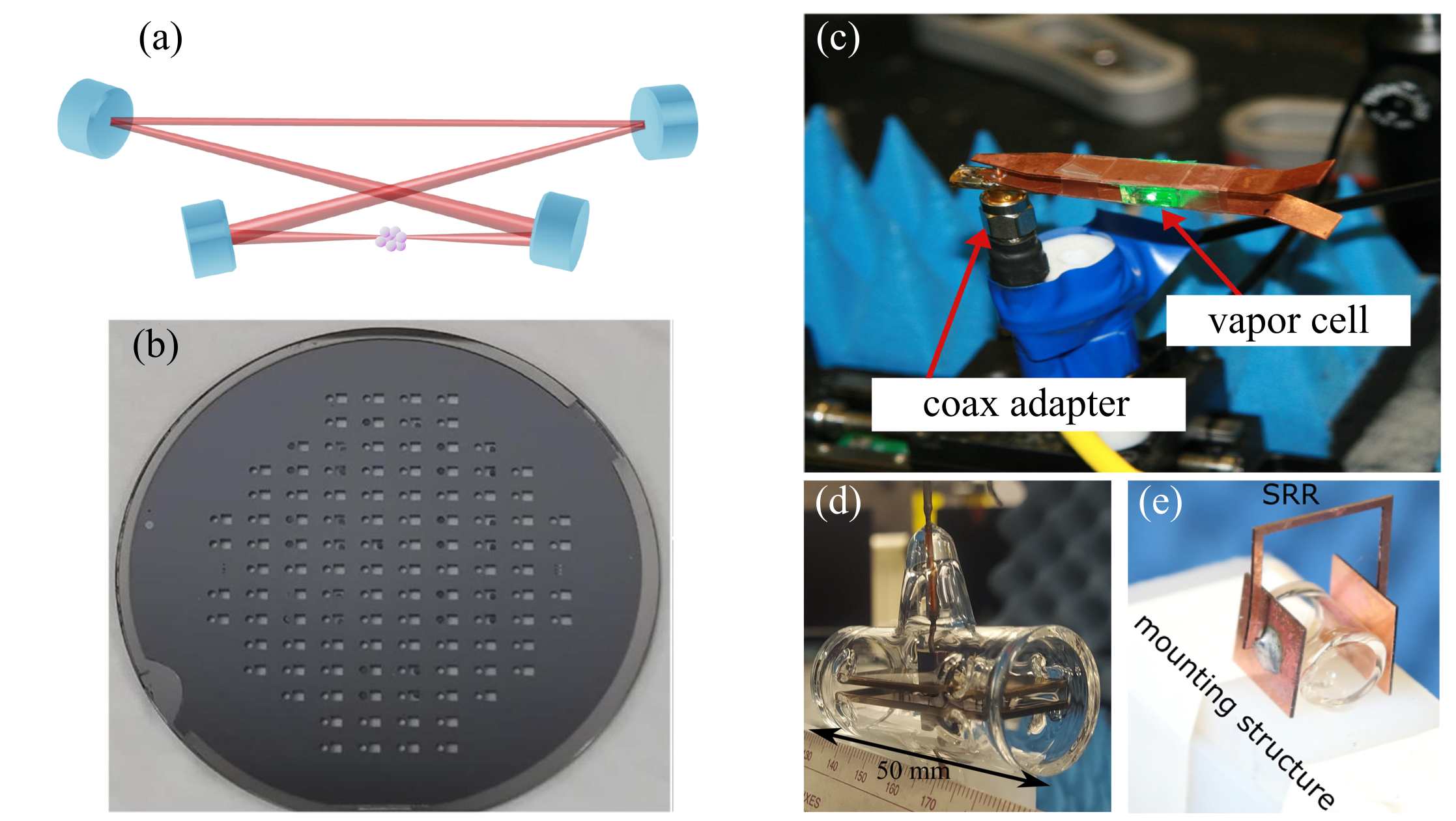}
	\centering
	\caption{Schematic diagrams of resonance cavities. (a) 4-mirror ring optical resonator \cite{Li2022}, (b) Micro-electro-mechanical systems (MEMs) vapor cell \cite{Zhu2023}, (c) parallel-plate waveguide (PPWG) \cite{Simons2019b}, (d) integrated PPWG \cite{Holloway2022a} and (e) split-ring resonator (SRR) \cite{Holloway2022}.}
	\label{fig9}
\end{figure}

The introduction of an optical/microwave cavity into typical measurement setup can enhance the intra-cavity circulating power of optical or microwave field, and extend the interaction length or amplify the microwave strength through multiple oscillations. Also, the intra-cavity atomic medium reduces absorption and generates large dispersion, resulting in narrowing of spectral line-widths. The common forms of resonance cavities are shown in figure\ref{fig9}.

The theoretical model for the coupling between cavity and Rydberg atoms was proposed, as well as phenomena in the coupling process like intra-cavity anomalous dispersion and collective Rabi Splitting were analyzed \cite{Peng2018,Yang2020}. Recently, an experimental attempt for enhancing Rydberg atoms based MW E-field sensing system by a 4-mirror ring optical cavity in coherent Rydberg $^{87}$Rb atoms was realized \cite{Li2022}. The novel cavity-atom coupling scheme shows a two-time improvement in measurement sensitivity has been achieved owing to an enhanced intra-cavity photon–atom coupling. Furthermore, integrating the vapor cell and the optical cavity with micro-electro-mechanical systems (MEMs) can go a step further to miniaturize the setup and provide key technical support for practical device \cite{Zhu2023,Noaman2023}.

The integration effect of microwave resonator improves the target E-field strength inside the cavity at the same input power, which means a higher measurement sensitivity \cite{Anderson2018}. When embedding the Rydberg atoms based sensor into a parallel-plate waveguide (PPWG), the novel configuration shows the capacity for amplitude and phase detection of a RF E-field \cite{Simons2019b}. The PPWG antenna here firstly radiates a LO field, the more important function is that it can capture the RF field arriving from a remote location and concentrate it at the location of atomic vapor cell for detection. The embedding of two stainless-steel parallel plates inside the vapor cell allows the measurement of DC and 60 Hz AC voltages with more favorable size, weight, power, and cost \cite{Holloway2022a}. Recently, the use of a sub-wavelength split-ring resonator (SRR) incorporated with an atomic vapor cell provides a 100 times enhancement of the E-field measurement sensitivity. The sensitivity of 55 nV$\cdot$cm$^{-1}$Hz$^{-1/2}$ was achieved by combining EIT with a heterodyne Rydberg atoms based mixer approach \cite{Holloway2022}.

2) Photon conversion from microwave to optical bands.

\begin{figure}[ht!]
	\includegraphics[width=15.0cm]{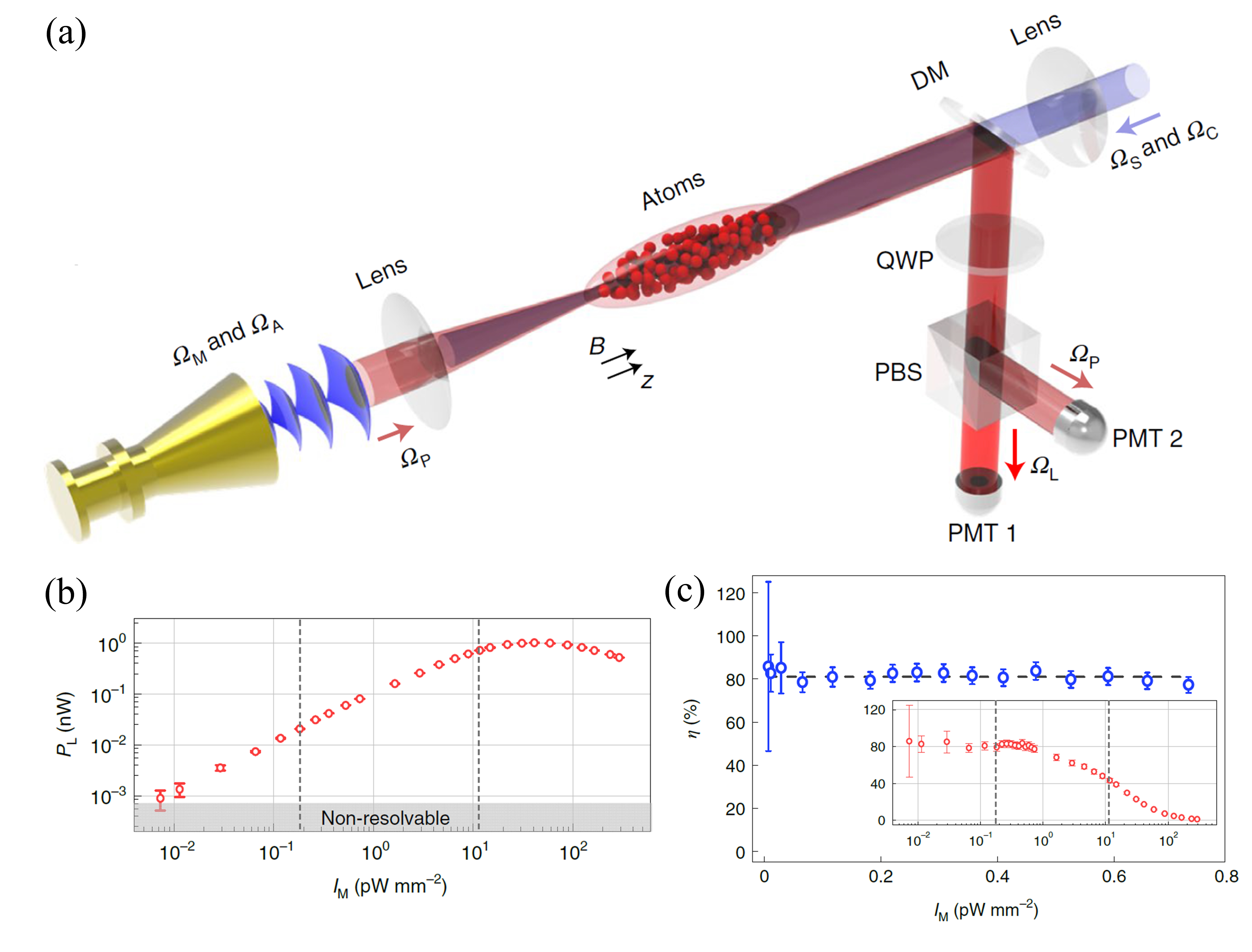}
	\centering
	\caption{Schematic diagrams and related results of off-resonant six-wave mixing cold atom converter \cite{Tu2022}. (a) Scheme of experimental setup. (b) Light power versus input microwave intensity. (c) Calculated conversion efficiency for a linear conversion region. The inset of (c) depicts conversion efficiency against the entire data of input microwave intensity in (b).}
	\label{fig10}
\end{figure}

Except for EIT and AT splitting mechanism, the other efficient optical readout scheme is a multi-wave mixing scheme converting a microwave field to an optical field. The conversion process satisfying energy and momentum conservation conditions permits the readout of amplitude and phase information of the microwave field from the optical field. The feasibility of scheme was firstly demonstrated in a cold rubidium atom with a photon-conversion efficiency of $\sim$0.3$\%$ at equivalent electric field strength of 6.56 mV$\cdot$cm$^{-1}$ and a broad conversion bandwidth of more than 4 MHz. \cite{Han2018}. With the continuous optimization of experimental conditions of vector matching, optical thickness and laser detuning, the conversion efficiency was improved to $\sim$5$\%$ \cite{Vogt2018,Vogt2019}. The coherent microwave to optics conversion using Rydberg atoms and off-resonant scattering technique with an efficiency of 82$\pm$2$\%$ and a bandwidth of $\sim$1 MHz was recently achieved, which breaks 50$\%$ threshold of the no-cloning quantum regime \cite{Tu2022}. Figure \ref{fig10} shows the schematic diagram and related results of the off-resonant six-wave mixing microwave to optical conversion. The joint use of photon counting and auto-correlation measurement in a free-space six-wave mixing process achieved the all-optical readout of photons in free-space 300 K thermal background radiation at 1.59 nV$\cdot$cm$^{-1}$Hz$^{-1/2}$ with the sensitivity down to 3.8 K of noise-equivalent temperature \cite{Borowka2023}. The combining of a three-dimensional microwave resonator, a vibration-stabilized optical cavity and a closed-loop four-wave-mixing process in a measurement system reached an internal conversion efficiency of 58(11)$\%$ with a conversion bandwidth of 360(20) kHz \cite{Kumar2023}. The extensions of this technique will allow near-unity efficiency conversion in both the mm-wave and microwave regimes.

3) Upgrading laser/microwave source.

The replacement of laser/microwave source in typical Rydberg atoms based MW E-field sensing system will bring improvement in the phase resolution and operating/instantaneous bandwidth.

Once can upgrade the single frequency coupling laser into a laser with symmetrical frequency side-bands generated by an electro-optic modulator, where the side-band frequency coincidentally corresponds to a pair of Rydberg states near-resonantly coupled by the same MW field. This allows to build an internal-state Rydberg atom interferometer. The configuration enables LO-free full 360$^{\circ}$-resolved phase measurement, and realized a phase resolution of 2 mrad \cite{Anderson2022,Berweger2022}. 

\begin{figure}[ht!]
	\includegraphics[width=15.0cm]{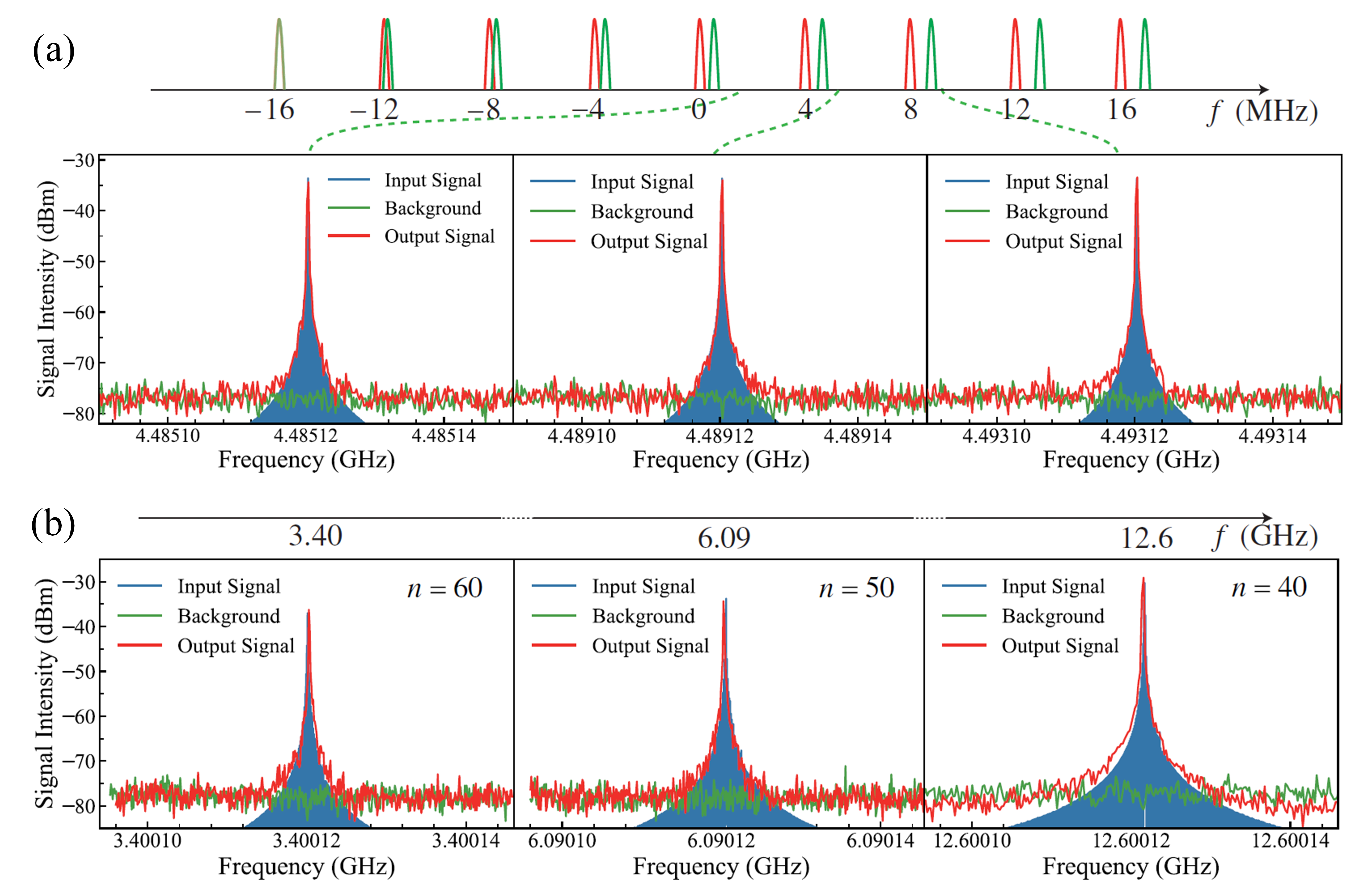}
	\centering
	\caption{Microwave frequency comb spectrum \cite{Zhang2022}. (a) Detected MW signals from different comb teeth. (b) Detected MW signals in different Rydberg states. }
	\label{fig11}
\end{figure}

The optical frequency comb-like (OFC-like) laser/microwave source includes multiple side-bands, which are capable to excite several Rydberg transitions independently and simultaneously. A Rydberg atomic sensor with OFC-like MW field provides quasi-continuous frequency readout of both the strength and absolute frequency of microwaves with operating bandwidth of 125 MHz and maximum instantaneous bandwidth of 300 kHz \cite{Zhang2022}. Figure \ref{fig11} shows a MW frequency comb spectrum. Also, Rydberg atoms based MW E-field sensing system with OFC-like probe laser was demonstrated, providing scan-free readout scheme with minimum detectable MW E-field strength of 66 $\mu$Vcm$^{-1}$ and sensitivity of 2.3 $\mu$V$\cdot$cm$^{-1}$Hz$^{-1/2}$ \cite{Dixon2022}.

4) The introduction of algorithm.

\begin{figure}[ht!]
	\includegraphics[width=15.0cm]{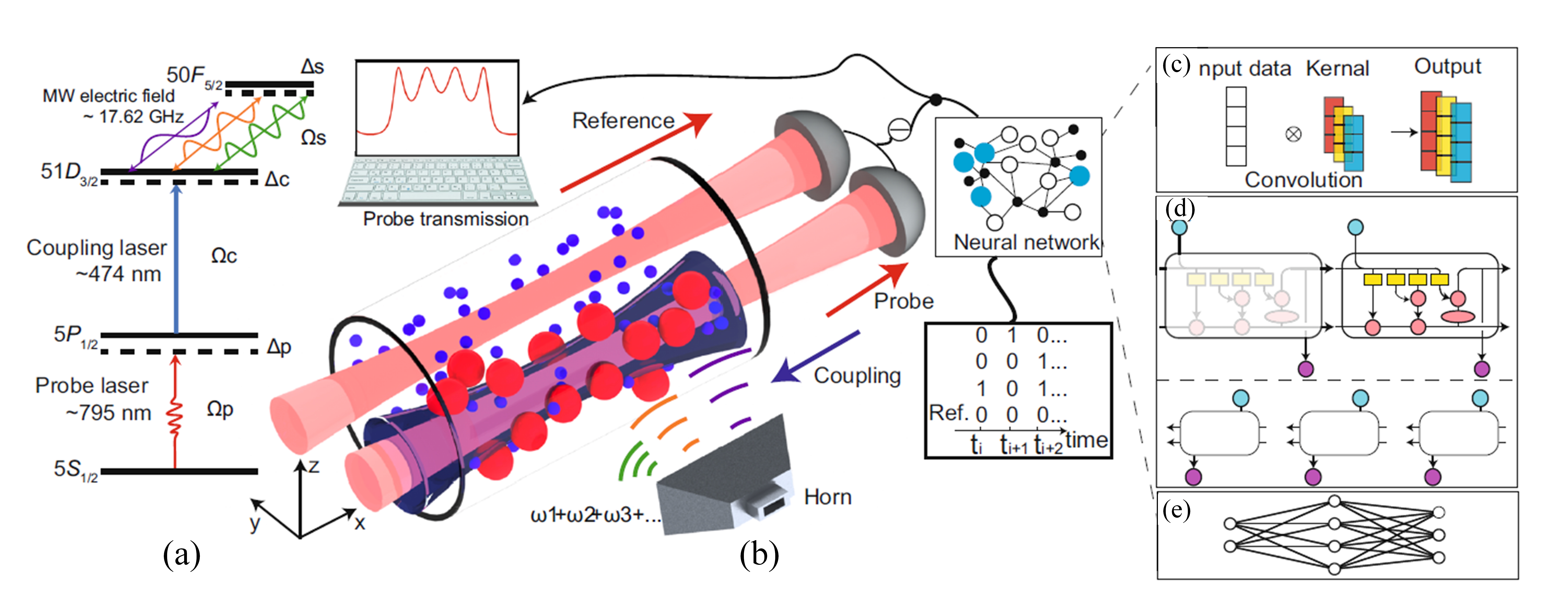}
	\centering
	\caption{Illustration of the setup \cite{Liu2022}. (a) Experimental energy diagram. (b) Schematic of a Rydberg atoms based MW receiver and mixer interacting with multi-frequency signals. (c–e) Schematics of the neural network. The network consists of (c) a one-dimensional convolution layer, (d) a bi-directional long-short term memory layer, and (e) a dense layer.}
	\label{fig13}
\end{figure}

The Rydberg atomic microwave sensing system shows no difference in sensitivity to target MW E-field and background electromagnetic noise, which is unfavorable for higher sensitivity. While with the support of cutting-edge artificial intelligence, the Rydberg atomic microwave sensing system has gained greater vitality. By combining Rydberg atoms with deep learning model, the direct receiving and decoding of the frequency-division multiplexed signal was achieved at high speed and with high accuracy \cite{Liu2022}. The illustration of scheme and method are shown in figure \ref{fig13}.

\subsection{Alternative Rydberg measurement medium} 

Compared with “warm atoms”, cold Rydberg atoms have the lower thermal velocity and can greatly reduce the collision probability among atoms. As the result, the coherence lifetime of cold atoms in high-$n$ Rydberg states increases to a level approaching its lifetime \cite{Fancher2021}. Using similar energy level configuration to the “warm atoms”, MW E-field strength was measured in an ultracold Rydberg atom ensemble bases on EIA spectra. The narrower line-width of EIA spectrum enables the minimum detectable electric field strength of $\sim$100 $\mu$V$\cdot$cm$^{-1}$, about 1/50 the low bound achievable by vapor-cell EIT method \cite{Liao2020}.

What's more worth mentioning, the transition between adjacent Rydberg circular states has the largest transition dipole moment, and the Rydberg circular state has the longest radiation lifetime under the same principal quantum number \cite{Raimond2001}. Recently, it is shown experimentally that a Rydberg circular state $^{85}$Rb atoms system with $n$=60 obtained lifetime of more than one millisecond at room temperature \cite{Wu2023}. Therefore, based on Rydberg circular state system, the sensitivity approaching projection noise limit allowed by the atomic system is predictable.

The samples of atomic vapor need containers for storage, while the containers are always fragile and require a suitable operating environment. The solid-state Rydberg excitons as more compatible  mediums have been gradually developed for MW E-field sensing. Among them, the metallic oxides are pioneers in this regard. The coupling between Rydberg excitons in cuprous oxide and MW E-fields using one-photon and two-photon spectroscopy techniques were performed with E-field strengths of 4 V$\cdot$cm$^{-1}$ and 2 V$\cdot$cm$^{-1}$, respectively \cite{Gallagher2022}.

\subsection{Quantum-enhanced measurement system} 
\label{5.3}
Although the measurement sensitivity of a Rydberg atomic MW E-field sensing system has been reported in recent years to gradually approach the shot noise limit, sensitivity cannot be pushed indefinitely by conventional experimental schemes. Breaking through the standard quantum limit requires the introduction of quantum enhanced measurement systems such as Schrödinger-cat state, interacting many-body state, and squeezed state.

The measurement of an E-field with an electrometer consisting of a large angular momentum carried by a single atom in a high energy Rydberg state was performed and the fundamental Heisenberg limit was approached when the Rydberg atom undergoes a non-classical evolution through Schrödinger-cat states. The method gives a single-shot sensitivity of 1.2 mV$\cdot$cm$^{-1}$ for a 100 ns interaction time \cite{Facon2016}. The many-body critical enhanced metrology for the sensing MW E-fields with a non-equilibrium Rydberg atomic system shows the enhanced sensitivity by three orders of magnitude increment compared with single-particle systems. The key transition points were recorded with the equivalent sensitivity to be 49 nV$\cdot$cm$^{-1}$Hz$^{-1/2}$ \cite{Ding2022}. 

Taking advantage of non-classical state system offers an attractive alternative to boost measurement sensitivity beyond the classical limit, which is a mountain that cannot be surmounted with classical coherent optical fields. The quantum enhanced two-photon Raman absorption spectroscopy of 5D$_{3/2}$ state of $^{87}$Rb with detection noises 5 dB below the classical shot noise limit was performed using a source of two-mode intensity squeezed light \cite{Prajapati2021a}.

\section{Outlook}
\label{soutlook}
The Rydberg atoms based MW E-field sensing system has shown the advantages of high sensitivity, broad bandwidth, stealth, traceability and integration in MW E-field sensing, but there are still many challenges to be solved. The aspects include but are not limited to the following: 1) Continue to pushing the boundaries of sensitivity and bandwidth. The joint use of superheterodyne scheme and quantum enhanced measuring system may achieve a new breakthrough in sensitivity. The upgrading of laser/microwave source and the multiplexing of atomic species seem good pathways to widen the operating and instantaneous bandwidths. 2) Access to existing antenna systems. The traditional antennae and their supporting equipment have been highly developed. It is necessary for Rydberg atoms based MW E-field sensing system to adapt for the traditional antenna systems, so that it can be connected and integrated to the existing antenna system and work in parallel. 3) Integrated devices working in non-laboratory setting. Most functions of existing Rydberg atoms based MW E-field sensing system are demonstrated in the laboratory environment, while most of applications for MW receiver are in the outdoor or even outer space environment. The task of MW E-field imaging and reconstruction also calls for such an integrated sensor as the technological path in microwave magnetic field sensing \cite{Horsley2015,Horsley2016}. More devices, for instance mature lasers with portability and stability, vapor cells with micro-integration, and fiber-coupled devices are the contents that need to be explored. 4) New data processing method based on advanced algorithms. The general Rydberg atoms based MW E-field sensing system requires massive operations during signal processing, which is time-consuming and labor-intensive. In the future, remote sensing and communication call for signal collection and processing systems to deal with difficulties like long operating time, the extraction of useful information in complex electromagnetic environment, multiple target frequencies and high response speed. Therefore, it is urgent to develop algorithms, such as deep learning and intelligent retrieval, and perform high-speed calculations with the obtained large quantity of data. 5) Distributed sensing of Rydberg atoms based MW receiver. During the measurement of multiple parameters distributed spatially, distributed quantum sensing is an important means for performing precise measurement tasks on multiple nodes in remote space simultaneously \cite{Liu2021}. Arranging Rydberg atoms based MW E-field sensors into arrays may become an effective solution, promising in resource utilization and the number of measurable parameters. Tackling these challenges will expand the existing applications as well as seek new applications of Rydberg atoms based MW E-field sensing technology in different extreme environments.

\section{Data availability statement}

No new data were created or analysed in this study. 

\section{Acknowledgments}

We acknowledge funding from the National Key R$\&$D Program of China (2022YFA1404003), Innovation Program for Quantum Science and Technology (Grant No.2021ZD0302100), the National Natural Science Foundation of China (Grant Nos. 61827824, 62075121 and 61975104),

\section{References}

\bibliographystyle{iopart-num}
\bibliography{review.bib}

\end{document}